\documentclass{aa}

\usepackage{subcaption}
\usepackage{txfonts}
\usepackage{multirow}

\begin{document}

\title{On the estimation of stellar parameters with uncertainty prediction from Generative Artificial Neural Networks: application to Gaia RVS simulated spectra}
   
   \titlerunning{Stellar parameterization from Gaia RVS spectra using GANNs}

   \author{C. Dafonte\inst{1} \and D. Fustes\inst{1} \and M. Manteiga\inst{2} \and D. Garabato\inst{1} \and M. A. \'Alvarez\inst{1} \and A. Ulla\inst{3} \and C. Allende Prieto\inst{4,5}}
          
   \authorrunning{C. Dafonte et al.}
    
   \institute{Universidade da Coru\~na (UDC), Dept. de Tecnologías de la Información y las Comunicaciones, Elvi\~na, 15071 A Coru\~na, Spain\\
            \email{dafonte@udc.es, dfustes@udc.es, daniel.garabato@udc.es, marco.antonio.agonzalez@udc.es}
      \and
      Universidade da Coru\~na (UDC), Dept. de Ciencias de la Navegación y de la Tierra, Paseo de Ronda 51, 15011 A Coru\~na, Spain\\              
      \email{manteiga@udc.es}
      \and
      Universidade de Vigo (Uvigo), Dept. de Física Aplicada, Campus Lagoas-Marcosende, s/n, 36310 Vigo, Spain\\
      \email{ulla@uvigo.es}
      \and
      Instituto de Astrofísica de Canarias, 38200 La Laguna, Tenerife, Spain
      \and
      Universidad de La Laguna, Departamento de Astrofísica, 38206 La Laguna, Tenerife, Spain\\
      \email{callende@iac.es}
           }

   \date{\today}

\abstract
   {}
   {We present an innovative artificial neural network (ANN) architecture, called Generative ANN (GANN), that computes the forward model, that is it learns the function that relates the unknown outputs (stellar atmospheric parameters, in this case) to the given inputs (spectra). Such a model can be integrated in a Bayesian framework to estimate the posterior distribution of the outputs.}
   {The architecture of the GANN follows the same scheme as a normal ANN, but with the inputs and outputs inverted. We train the network with the set of atmospheric parameters ($T_{eff}$, $log\:g$, $\lbrack Fe/H \rbrack$ and $\lbrack \alpha/Fe \rbrack$), obtaining the stellar spectra for such inputs. The residuals between the spectra in the grid and the estimated spectra are minimized using a validation dataset to keep solutions as general as possible.}
   {The performance of both conventional ANNs and GANNs to estimate the stellar parameters as a function of the star brightness is presented and compared for different Galactic populations. GANNs provide significantly improved parameterizations for early and intermediate spectral types with rich and intermediate metallicities. The behaviour of both algorithms is very similar for our sample of late-type stars, obtaining residuals in the derivation of $\lbrack Fe/H \rbrack$ and $\lbrack \alpha/Fe \rbrack$ below $0.1$ dex for stars with Gaia magnitude $G_{rvs}<12$, which accounts for a number in the order of four million stars to be observed by the Radial Velocity Spectrograph of the Gaia satellite.}
   {Uncertainty estimation of computed astrophysical parameters is crucial for the validation of the parameterization itself and for the subsequent exploitation by the astronomical community. GANNs produce not only the parameters for a given spectrum, but a goodness-of-fit between the observed spectrum and the predicted one for a given set of parameters. Moreover, they allow us to obtain the full posterior distribution over the astrophysical parameters space once a noise model is assumed. This can be used for novelty detection and quality assessment.}

   \maketitle

\section{Introduction\label{sect:intro}}

The first automatic systems for spectral classification were developed in the 80s and based on two paradigms: expert systems and pattern recognition. Knowledge-based systems were created by defining a set of rules/models that relate, for the case of astronomical objects, spectral indexes (absorption line equivalent widths, colour indexes, etc.) with a spectral class in the MK classification system \citep{1943assw.book.....M}. Examples of these types of systems can be found in \cite{1981AJ.....86.1360T} and \cite{1982BICDS..23...39M}. Systems based on pattern recognition rely on a distance function (cross-correlation, euclidean, chi-squared) that is minimized between the observed spectra and a set of templates, so that the observed spectrum receives the class of the closest template. In the 90s, machine learning methods, specially artificial neural networks (ANNs), began to be applied for MK classification (see the works from \citealt{1994MNRAS.269...97V} and \citealt{1995ApJ...446..300W}). The use of ANNs offered several advantages. Firstly, it is not necessary to explicitly define spectral indexes and models to obtain the classification. Furthermore, once the ANN has been trained, its application is really fast in comparison with distance minimization schemes. Finally, ANNs provide accurate results even when the signal to noise ratio (S/N) of the spectra is very low. This property represents an important advantage in the analysis of extensive surveys, with a high percentage of low S/N data, as is the case with the Gaia survey.

By the end of the 90s, a movement lead by researchers such as \cite{1997MNRAS.292..157B} and \cite{2001ApJ...562..528S}, changed the perspective of stellar spectral classification towards a process of astrophysical parameter (AP) estimation, which is a problem that resembles the nonlinear regression problem in statistics. ANNs are known to perform very well in nonlinear regression regimes, so they have remained in the state of the art for AP estimation. During the first decade of the 21st century, researchers such as \cite{2010PASP..122..608M} proposed the combination of ANNs and wavelets for improving the estimations as a function of the spectral S/N.

One of the main criticisms of ANNs, as well of other machine learning schemes, is that they are incapable of providing an uncertainty measure on their solutions. Some authors have proposed schemes to provide confidence intervals in addition to the ANN outputs. To do this, it is necessary to take into account the different sources of error such as the training data density, target intrinsic noise, ANN bias, error in the observations acquisition, and the mismatch between training data and observations. Furthermore, such errors can be input dependent. For example, \cite{710658} presents an approximated Bayesian framework that computes the uncertainty of the trained weights, that can then be used to obtain the uncertainty predictions taking into account several sources of error that are input dependent. However, such a method requires the computation of the Hessian matrix, which is not feasible for large networks. The ANNs needed for AP estimation are usually very large because the network inputs are as many as the number of spectrum pixels. This number depends on the wavelength coverage and the spectral resolution, but usually is of the order of thousands. Other methods, such as bootstrapping, also require a large number of computations, which makes them unfeasible for large problems.

Gaia, the astrometric cornerstone mission of the European Space Agency (ESA) was successfully launched and set into orbit in December 2013. In June 2014, it started its routine operations phase scanning the sky with the different instruments on board. Gaia was designed to measure positions, parallaxes and motions to the microarcsec level, thus providing the first highly accurate 6-D map of about a thousand million objects of the Milky Way. Extensive reviews of Gaia instruments modes of operation, the astrophysical main objectives and pre-launch expected scientific performance, can be found, for example, in \cite{2013hsa7.conf...82T}.

A vast community of astronomers are looking forward to the delivery of the first non-biased survey of the entire sky down to magnitude 20. Moreover, the final catalogue, containing the observations and some basic data analysis, will be opened to the general astrophysical community as soon as it is be produced and validated. The definitive Gaia data release is expected in 2022-2023, with some intermediate public releases starting around mid-2016 with preliminary astrometry and integrated photometry. The Gaia Data Processing and Analysis Consortium (DPAC) is the scientific network devoted to processing and analysing the mission data. The coordination unit in charge of the overall classification of the bulk of observed astronomical sources by means of both supervised and unsupervised algorithms is known as CU8. This unit also aims to produce an outline of their main astrophysical parameters. The CU8 Astrophysical Parameters Inference System (APSIS, \citealp{2013A&A...559A..74B}) is subdivided into several working packages, with GSP-Spec (General Stellar Parameterizer - Spectroscopy) being the one devoted to the derivation of stellar atmospheric parameters from Gaia spectroscopic data.

GSP-Spec will analyse the spectra obtained with Gaia Radial Velocity Spectrograph (RVS) instrument. Though its main purpose is to measure the radial velocity of stars in the near infrared CaII spectral region, it will also be used to estimate the main stellar APs: effective temperature ($T_{eff}$), logarithm of surface gravity ($log\:g$), abundance of metal elements with respect to hydrogen ($\lbrack Fe/H \rbrack$) and abundance of alpha elements with respect to iron ($\lbrack \alpha/Fe \rbrack$). The software package being developed by the GSP-Spec team is composed of several modules which address the problem of parameterization from different perspectives \citep{2006MNRAS.370..141R, 2010PASP..122..608M}, and has been recently described in \cite{2016A&A...585A..93R} (from now on, RB2016). This work focuses on developments carried out in the framework of one of these modules, called \textit{ANN}, that is based on the application of ANNs.

\begin{figure}[h]
  \centering
    \includegraphics[width=0.75\columnwidth]{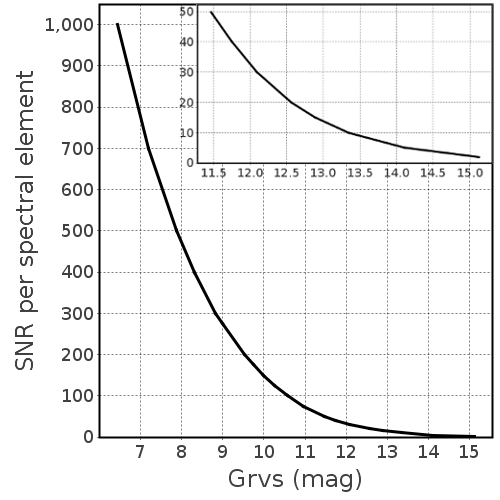}
  \caption{Signal to noise ratio as a function of the star magnitude $G_{rvs}$ for RVS post-launch configuration \citetext{D. Katz, 2015, private communication}}
  \label{fig:SNRvsGrvs}
\end{figure}

During the commissioning stage of the mission (from February to June, 2014) unexpected problems were found that lead to a degradation of RVS limiting magnitude to a value close to $G_{rvs}=15.5$ mag \citep{2014EAS....67...69C}, that is around 1.5 mag brighter than expected. Figure \ref{fig:SNRvsGrvs} shows updated end-of-mission values for the $G_{rvs}$ versus S/N relationship for resolution element (3 pixels) that are based on simulations of RVS post-launch performance. The different algorithms for RVS stellar parameterization developed in the framework of Gaia DPAC need to be evaluated by the use of synthetic spectra at a variety of S/N values, which correspond to different magnitude levels. These values, then, already incorporate the revised performance figures. From RB2016 it was clear that ANNs give in general better results at very low S/N, this is one of the motivations of studying in detail such an approach for stellar parameterization, and also addressing the problem of ANN uncertainty estimations with Generative ANNs (GANNs).

Uncertainty estimation of computed APs is crucial for the validation of the parameterization itself and for the exploitation of the results by the astronomical community. Therefore, some of the algorithms being developed in Gaia DPAC have addressed this problem. The idea is to change the perspective of the regression problem by learning the forward model (also called generative model) instead of the inverse model, which then allows the comparison between the observed spectrum and the spectrum estimated by the generative model. One proposed algorithm is Aeneas \citep{2012MNRAS.426.2463L}, that has been integrated in CU8 software chain APSIS. Aeneas defines a generative model that predicts the spectra from a set of APs by means of modelling with splines Gaia spectrophotometers data. Then, for a given spectrum, it finds the set of parameters that provide the estimated spectrum which maximizes the likelihood with respect to the observed one. To do so, it uses Markov chain Monte Carlo (MCMC, \citealt{10.2307/2346063}) algorithms to search for the best APs. The generative model is integrated in a Bayesian framework that enables the computation of the posterior distribution of the parameters given the observed spectrum. In this work, we also present a generative model but now based on neural networks, Generative ANNs, for AP estimation from Gaia RVS spectra. We discuss its performance in comparison with a classical ANN feed-forward algorithm.

The remainder of this paper is organized as follows: in Section \ref{sect:simulations} we describe the library of synthetic spectra that is being used to test the algorithms, in Sections \ref{sect:anns_algo}, \ref{sect:ganns_algo} and \ref{sect:implem} we describe the ANN and GANN algorithms and their performances and in Section \ref{sect:results} we show the results obtained when it is applied to the Gaia RVS simulated data. Finally, in Section \ref{sect:discussion} we discuss the advantages and drawbacks of the proposed method.

\section{Simulation of RVS spectra\label{sect:simulations}}

The Gaia RVS instrument is currently obtaining spectra between 847 and 871 nm for relatively bright stars among the ones observed by Gaia, with magnitudes in the range $6<G_{rvs}<15.5$. To mitigate post-launch performance degradation, the instrument is operating only in high resolution mode, with $R=11200$ (see for instance \citealp{2009sf2a.conf...57K}). In RB2016 we discussed the internal errors that can be expected in the derived stellar parameters as a function of the stellar brightness, considering simulations of RVS data and different parameterization codes developed within the GSP-Spec working group. In this paper we introduce a novel approach that can extend the capabilities of neural networks for parameterization problems.

A library of spectra was generated and it covers all the space of APs to be estimated. The library contains simulations for stars of spectral types from early B to K, generated by means of \cite{2003IAUS..210P.A20C} grid of stellar atmospheric models. A similar library is described in detail in RB2016. Our library has been generated with the same model atmospheres and spectral synthesis code than in RB2016. It is composed by two grids, an `early to intermediate type stars' sample and an `intermediate to late type stars' sample (from now on, early stars and late stars samples). Their coverage and resolution are shown in Table \ref{tab:grids}. The $[\alpha/Fe]$ parameter was not estimated for the sample of earlier type stars, since they barely show absorption lines related to metallic elements. We shall refer to this library as the nominal grid of synthetic RVS spectra.

\begin{table}[h]
\begin{center}
\begin{tabular}{llll} \hline
\textbf{Class} &\textbf{AP} &\textbf{Range} &\textbf{Resolution}\\ \hline
\multirow{3}{*}{Early} & $T_{eff}$ & [7000,11500] & 500K \\
 & $log\:g$ & [2,5] & 0.5dex \\
 & $[Fe/H]$ & [-2.5,0.5] & 0.5dex \\ \hline
\multirow{4}{*}{Late} & $T_{eff}$ & [4000,8000] & 250K \\
 & $log\:g$ & [2,5]& 0.5dex \\
 & $[Fe/H]$ & [-2.5,0.5]& 0.5dex \\
 & [$\alpha/Fe$] & [-0.4,0.8] & 0.2dex \\ \hline
\end{tabular}
\end{center}
\caption{Library of simulated spectra for training RVS parameterization algorithms. The range of values and resolution of the different APs is shown for both early stars and late stars 
samples, as defined in the text.}
\label{tab:grids}
\end{table}

We have considered a RVS noise model based on updated instrument performance information available for DPAC. As detailed in RB2016, the noise properties depend largely on the star brightness. We account for Poisson shot noise in the data and for the charged coupled device (CCD) read-out-noise, which is assumed to be $4e-$. Since the final spectra will be accumulated from a number of epochs, 100 visits were assumed, and since objects typically cross three CCDs per visit, we simulated individual observations (spectra acquired per CCD per visit), and then combined them to produce an end-of-mission $1-\sigma$ noise spectrum for each source. Mostly, Gaussian read-out-noise dominates and it represents a good approximation for the overall noise behaviour. Figure \ref{fig:SNRvsGrvs} shows the relationship between the star magnitude $G_{rvs}$ and the S/N of the RVS spectrum for post-launch instrument configuration (David Katz, 2015, private communication).

Taking into account the discussion about the ANN algorithm performance in RB2016, we considered that it could be worthwhile to better focus our study on low S/N spectra. With this aim, we computed noised versions of our nominal grid at six levels of S/N: $356$, $150$, $49$, $13.8$, $5.7$ and $2.4$,  which correspond to $G_{rvs}$: $8.5$, $10$, $11.5$, $13$, $14$ and $15$, respectively. From these nominal grids, interpolations at random combinations of the four atmospheric parameters were performed, obtaining a total of 20.000 random spectra at each selected $G_{rvs}$ magnitude. Finally, a subsample of 10400 spectra was selected from the random samples, combining the atmospheric parameters according to reasonable limits for the ages of stars populating the Milky Way, following the same procedure explained in RB2016. Since our tests were conducted only in high resolution mode, we had to re-run the simulations of both nominal and random datasets, so we can not guarantee that our random dataset is exactly the same as the one used in RB2016 due to its random nature, although it was filtered using the same isochrones. During the training phase, the nominal dataset will be used to train both ANNs and GANNs, and a subset of 100 random spectra from the random dataset will be used for validating the networks. Once this stage is completed, the rest of the random dataset will be presented to the ANNs and GANNs in order to evaluate their performances and compare them (see Section \ref{sect:results}).

Typical high resolution RVS spectra are shown in RB2016, illustrating the features present for a variety of stellar types. Late and intermediate type spectra are more sensitive to temperature and metallicity, while spectra from the hot stars are dominated by the star gravity. Obviously, it is expected that the estimation of stronger APs was more robust against noise than the estimation of weakest ones.

\section{ANNs for stellar parameterization\label{sect:anns_algo}} 

\begin{figure}[h]
  \centering
  \includegraphics[width=.9\columnwidth]{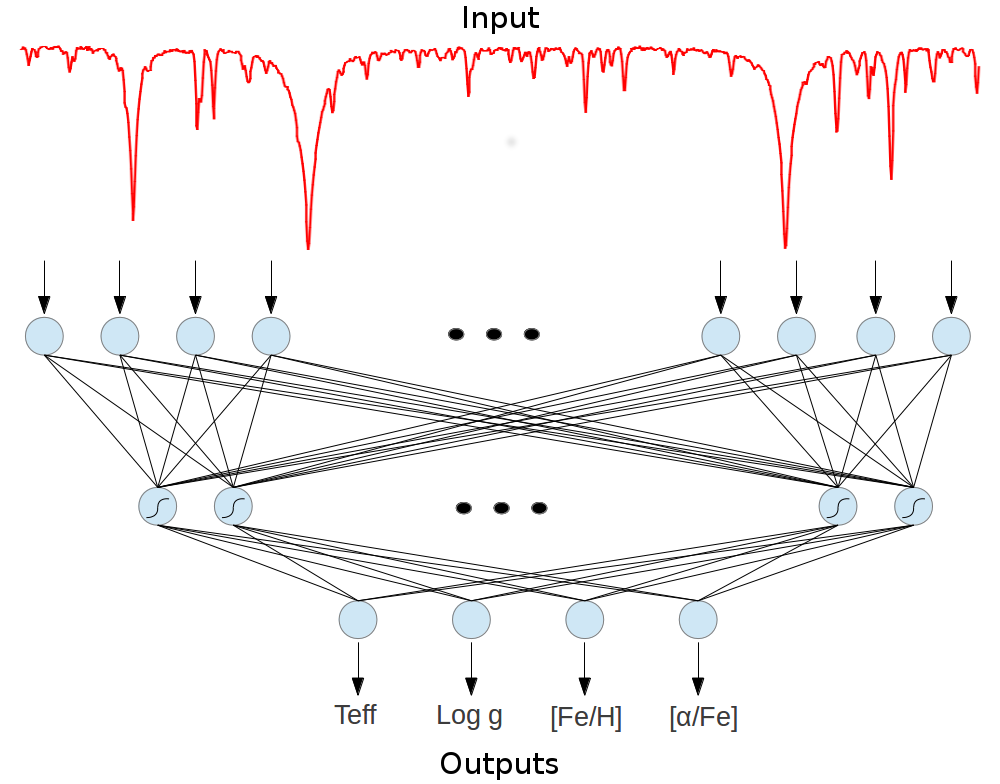} 
  \caption{ANN architecture for AP estimation.}
  \label{fig:ANN}
\end{figure}

The ANN model, designed to be used for AP estimation in GSP-Spec, is a three-layered fully-connected feed-forward network, with as many inputs as pixels in the spectrum and as many outputs as the number of parameters to be estimated, following the scheme shown in Figure \ref{fig:ANN}.

The neurons in the input and output layers have a linear activation function:
\begin{equation}\label{eq:linear_act_fun}
  f(x) = x
,\end{equation}
while those in the the hidden layer have a logistic activation function:
\begin{equation}\label{eq:logistic_act_fun}
  f(x) =  \frac{1}{1 + e^{-x}}
.\end{equation} 

This architecture allows the neural network to approximate any nonlinear real function, provided that the weights were properly set and the hidden layer contains enough neurons. We trained the ANNs using an online backpropagation algorithm (generalized delta rule, GDR), which can be treated as a minimization problem:
\begin{equation}\label{eq:backprop_min}
  Min_W E
\end{equation}
where W is the configuration of the network (layers, number of hidden neurons, weights, etc.) and $E$ is an error function that evaluates the error between the outputs produced by the network and the desired ones:
\begin{equation}\label{eq:backprop_error}
  E = \frac{1}{P} \sum\limits_{p=1}^P E_p
,\end{equation}
where $P$ is the number of patterns in the training dataset, and $E_p$ is the error associated with the pattern $p$ from the training dataset:
\begin{equation}\label{eq:backprop_lse}
  E_p = \frac{1}{2}\sum\limits_{k=1}^K (d_{pk} - y_{pk})^2
,\end{equation}
where $K$ is the dimensionality of the output, and $d_p$ and $y_p$ are the desired output and the output obtained by the network for the pattern $p$ from the training dataset, respectively.

Additionally, an early stopping strategy was used to obtain the state of the network that best generalizes, that is, the one that minimizes the residuals between the desired and the obtained APs for the validation dataset.

Following the procedure described in \cite{2010PASP..122..608M} and \cite{Ordonez20101719}, either a wavelet transformation of the domain \citep{Meyer1989} or the plain flux vs wavelength was chosen to compute the artificial networks, since for low S/N values the wavelet filtered version of the inputs yields better parameterizations. This wavelet transformation was performed by means of Mallat decomposition \citep{192463} into three orders of `approaches' and `details'. Since they gave empirically better parameterizations, we only took into account the approaches.

The ANN configuration's weights will be randomly initialized within the range $[-0.2, 0.2]$, since it has been empirically proved that the networks offered better performances when the values of the initialization weights were limited. Hence, the only free network parameters that remain to be set, the learning rate ($lr$) and the number of hidden neurons ($nh$), are determined using particle swarm optimization algorithm (PSO, \citealp{488968}). PSO was initially intended for simulating social behaviour, and it works by having a population (called a swarm) of candidate solutions (called particles). These particles move around in the search-space according to simple mathematical constrains. When improved positions that minimize a cost function are discovered, these will guide the movements of the swarm. The process is repeated, until a satisfactory solution is eventually discovered. In the case of ANNs, PSO performs an efficient search for the best training parameters, those which minimize the residuals for the validation dataset (see Equation \ref{eq:backprop_min}, now applied to validation patterns instead of training patterns). We have empirically determined optimal values for $lr = 0.12$ and $nh = 60$.

Regarding the learning phase, we also need to establish the number of iterations that will be used to train the network. Since we are using an early stopping strategy, and therefore will obtain the network that best generalizes within the training phase, we have decided to iterate over $1000$ times, ordering the training dataset randomly, and checking the validation error for each $100$ iterations. Furthermore, to reinforce the generalization capabilities of the networks, we have decided to repeat the overall process ten times, obtaining ten independent ANNs, and then selected the ANN that obtained the smallest validation error to use it during the testing phase.

Furthermore, we conducted a study to check the influence of the random component of the ANNs training procedure: random weights initialization and the order of the training patterns. To this end we trained $20$ ANNs for early and late stars considering all the magnitudes, and measured the errors associated with both the validation and the testing datasets. In general terms, we found that the values of the mean and the standard deviation increased as the magnitude also increased, as well as for the early stars sample, less numerous than the late one. The use of an early stopping strategy in combination with several epochs of training, as well as the random ordering for the training dataset, allowed us to choose among the best networks. Obviously, such intrinsic uncertainties will be reflected in the mean errors and in the confidence intervals that we are reporting in Section \ref{sect:results}. It is also noticeable that the performance of these networks strongly depends on the training dataset and on its inherent quality, so a well-defined representative dataset must be used during this phase to ensure that the neural network learns the regression function and it can generalize properly. This fact can be guaranteed by the use of the nominal grid of spectra, which is calculated at regular intervals of stellar parameter space.

Section \ref{sect:results} shows the results obtained when the trained networks are applied to the test datasets. The performance of our ANN algorithm for RVS parameterization was discussed in RB2016. In that paper, this algorithm was trained and tested with high resolution simulated spectra for the brightest stars ($G_{rvs}<10$) and with low resolution spectra for the remaining dataset. The results obtained were subsequently extrapolated for the definitely adopted high resolution format. In this paper we take advantage of the possibility of performing the parameterization with the suited resolution for the complete range of magnitudes, to compare the results with those produced by the GANNs.

\section{GANNs for stellar parameterization\label{sect:ganns_algo}}

\begin{figure}[h]
  \centering
  \includegraphics[width=.9\columnwidth]{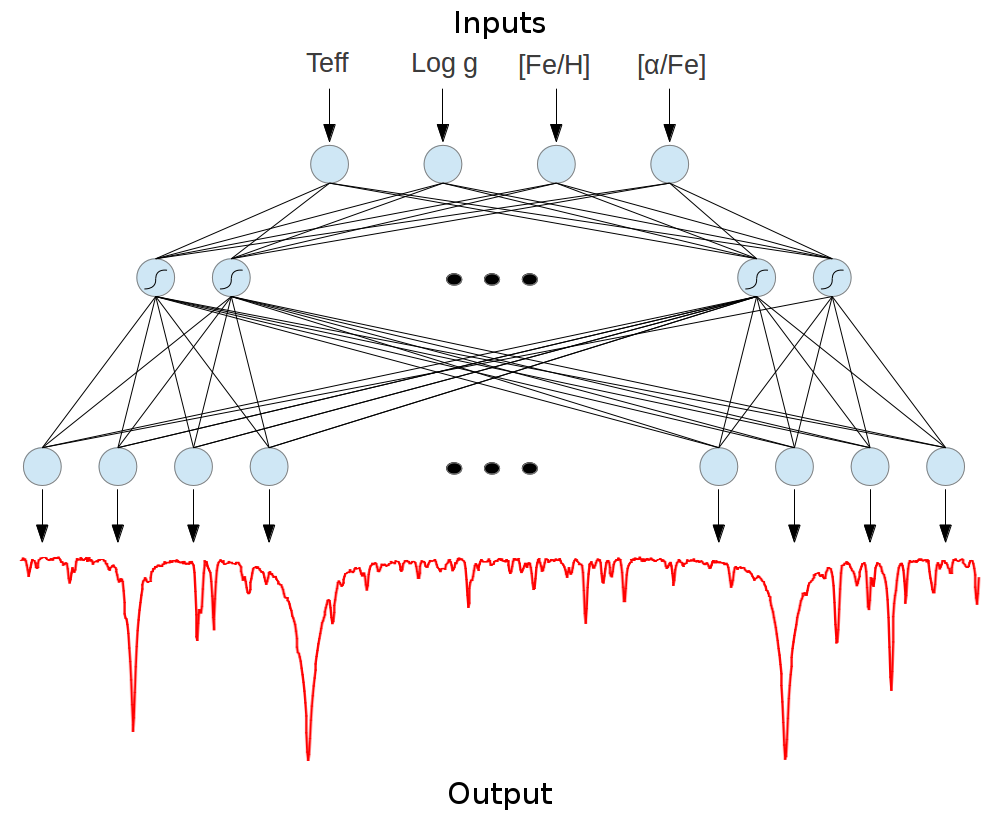} 
  \caption{GANN architecture for AP estimation.}
  \label{fig:GANN}
\end{figure}

Generative Artificial Neural Networks follow the same scheme as the normal ANNs, but with the inputs and the outputs inverted, as can be observed in Figure \ref{fig:GANN}. The training methodology for GANNs is also equivalent to the one used for ANNs, but in this case we try to minimize the residuals between the spectra in the grid and the estimated spectra, using again the validation dataset to keep general solutions.

In this field, uncertainty estimation in the outputs is nowadays a very active topic, as the involved calculations are both weighty and complex. There are several sources contributing to errors in ANN estimations. First of all, there is the error in the model produced by the lack of density in the training set, the error in the desired outputs $Y*$ and the one due to the flawed fitting of the neuron weights, $W$. These errors depend on the inputs $X$, and \cite{497809} proposed a way to calculate uncertainty estimations associated with them. They approached the problem by Bayesian inference, where, given the inputs, the neuron weight uncertainties can be estimated by the likelihood function $P(W|X)$. Then, the uncertainty in the networks output is also estimated depending on the inputs and the optimal weights $P(Y|X,W*)$. As the ANN is a nonlinear function, the uncertainty is approximated by a second order Taylor series, implying that the Hessian matrix has to be calculated. Other researchers have faced the problem in a different way, by extending the ANN architecture adding additional neurons in the hidden and output layers \citep{WeigendNix1994}. In our case, as is customary in experimental sciences, data come from instruments and sensors whose measurements are subject to errors that can be evaluated to some extent. As we mentioned before, previous works (for instance \citealp{710658}) have already incorporated those errors to the output uncertainty estimation using a Bayesian framework. For AP estimation using RVS spectra, we are dealing with networks containing $60,000$ weights, which translates to a $60,000* 60,000$ double precision floating points in the Hessian matrix. Such a calculation is technically unfeasible and requires memory storage of about 27 GB. Even the use of a Monte Carlo method to sample the inputs would be unworkable. GANNs can approach the uncertainty estimation using Bayesian inference from a different perspective. Generative models set out the direct problem, that is obtaining the observation from the APs to be estimated, instead of deriving the parameters from the observation. In this way, one can choose between estimating a set of optimal APs or finding out the parameters posterior probability distribution $P(AP)$, given the observed spectra $P(S)$. More specifically, once GANNs have been trained, they can be applied in two fashions:

\begin{enumerate}
 \item Maximum likelihood: If we are interested only in the best APs, we need to apply a procedure to find the parameters that maximize the likelihood (minimize the residual) between each observed spectrum and the spectrum estimated by the GANN (given the input set). This is achieved by an optimization procedure such as PSO, to search efficiently for the APs that maximize the likelihood (see Equation \ref{eq:likelihood}). In this sense, we can set the parameterizations calculated by normal ANNs or other approaches or methods, such as Aeneas, as an initial seed for the search, which reduce greatly the number of GANN estimations required to reach the optimum APs. This approach does not give AP uncertainties, but it still gives a goodness-of-fit measure and it only requires a few GANN evaluations to get the optimum parameters from the initial seed.
 
 \item Fully Bayesian: This approach allows us to obtain the full posterior distribution over the APs given the observed spectrum, $P(AP|S)$. This can be obtained following Bayes rule by:
 \begin{equation}\label{eq:bayes}
    P(AP|S)=\frac{P(S|AP)P(AP)}{P(S)}
 ,\end{equation}
 where $P(S|AP)$ is the likelihood of the estimated APs given the observed spectrum $S$, $P(AP)$ is the prior distribution over the APs and $P(S)$ is a normalization factor. The computation of the likelihood requires a noise model for the observed spectrum. In our case, it is assumed that the noise is Gaussian and independently distributed. Therefore the likelihood function is:
 \begin{equation}\label{eq:likelihood}
    P(S|AP)=e^{-d/2},\smallskip d=\sum_{i=1}^{N}\left(\frac{s_{i}-f_{i}(AP)}{\theta_{i}}\right)^2
 ,\end{equation}
 where $s_{i}$ is the observed flux in the band $i$ of the RVS spectrum, $\theta_{i}$ is the standard deviation of the Gaussian noise around the observed flux  and $f_{i}(AP)$ is the output $i$ of the GANN for a given $AP$ set. If the noise model is not Gaussian, then an appropriate likelihood function should be used instead. The prior distribution over the APs should cover the whole range. Due to the fact that our tests have been performed by the use of synthetic spectra, we have considered a uniform distribution for the priors over the APs, but any other distribution considered suitable for the application (for instance AP dependence according to evolutionary tracks) could be used instead. 
 \end{enumerate}

 The processing of the GANNs can be computationally hard, specially when a high number of APs is involved, since we need to evaluate them for a high number of combinations. Efficient sampling methods, such as Markov chain Monte Carlo (MCMC) methods, could help to reduce the number of evaluations in a future implementation.

GANNs are able to give not only the APs for a RVS spectrum, but a goodness-of-fit between the observed and the predicted spectrum for the given APs. This can be used for novelty detection and quality assessment in a project that involves the analysis of complex and very large databases like the Gaia survey. Additionally, if a Bayesian approach is adopted, the full posterior distribution over the APs can be obtained, which is a non-parametric measure of their uncertainty. 

\section{Implementation and computational efficiency\label{sect:implem}} 

Our stellar spectra parameterization software has been developed in Java, the programming language that was chosen by Gaia DPAC for the implementation of all Gaia processing and analysis working packages. Java allows all performance tests to be executed in the same homogeneous and stable platform. Our code for defining and training ANNs is integrated in a library named NeuralToolkit. Unit tests to check the right functioning of the ANN components together with visual utilities are included in this library. The PSO algorithm has been included in another library, OptimizationToolkit, while the facilities to handle RVS spectra as well as the derived atmospheric parameters are in the GSPSpecNNTests library, which, in fact, includes the two previous ones.

Particularly problematic is the evaluation of the likelihood function defined in Equation \ref{eq:likelihood} (Section \ref{sect:ganns_algo}) during PSO computation within the testing phase. The distance $d$, between observed and estimated spectra, can be as high as $10^{4}$ in the case of RVS spectra. This implies that when we calculate the negative exponential function of $d$, Java floating point is insufficient. To handle this problem, the complete set of distances $d$ are previously calculated, and subsequently they were normalized to the [0,1] interval.

\begin{table}[h]
\begin{center}
\begin{tabular}{ccp{.25\columnwidth}r}
\hline
\textbf{Net type} & \textbf{Domain} & \textbf{Training and validation time} & \textbf{Testing time}\\
\hline
\multirow{4}{*}{ANN} & lambda & \multicolumn{1}{r}{$10$m $48.3$s} & $2.2$s\\
 & A1\_db5 & \multicolumn{1}{r}{$5$min $48.2$s} & $1.1$s\\
 & A2\_db5 & \multicolumn{1}{r}{$2$min $53.4$s} & $0.5$s\\
 & A3\_db5 & \multicolumn{1}{r}{$1$min $28.2$s} & $0.2$s\\
\hline
GANN & lambda & \multicolumn{1}{r}{$10$min $39.6$s} & $15$min $6.2$s\\
\hline
\end{tabular}
\end{center}
\caption{Execution time measurement and comparison between both ANNs and GANNs for 10th magnitude early type stars during training and validation, and testing stages. Lambda domain refers to plain flux vs wavelength input data, while An\_db5 refers to the n-order wavelet approximation of the data. An \emph{Intel Core i7 950 (@ 3.07GHz \(\times\) 8)} with \emph{12GB DDR2 (2GB \(\times\) 6)} under \emph{Debian 8 Jessie - 3.16.0-4-amd64} was used.\label{tab:execution_times}}
\end{table}

To illustrate the computational efficiency of our algorithms, in Table \ref{tab:execution_times} we compare the computation times for training and validation, and test for ANNs and GANNs. In the case of wavelength domain, without dimensionality reduction, the computational times for training and validation are similar in both types of networks, but testing GANNs implies times that are a factor of 450 times higher due to the fact that the execution of the PSO calculations implies the evaluation of a high number of parameterizations. Times for the wavelet data domains used for computations are also shown.

\section{Results\label{sect:results}}

This section presents the results obtained by both ANNs and GANNs when are applied to the testing dataset described in Section \ref{sect:simulations}, once the networks are trained using the procedure described in Sections \ref{sect:anns_algo} and \ref{sect:ganns_algo}. As mentioned, in RB2016, the expected parameterization performances of different algorithms, including ANNs, are presented and discussed. The results for these networks presented in that paper were extrapolated for high resolution Gaia RVS spectra from parameterization experiments performed on both low and high resolution spectra. The new computations performed here allow us to check the trends and accuracy in the parameter estimation as well as to compare such results with the ones obtained for our GANN algorithm.

An exhaustive description of the parameterization accuracy for the complete mosaic of spectral types, metallicity cases and evolutionary stages in the Galaxy is beyond the scope of this paper. In the following we will comment on the accuracy trends of some relevant cases and compare the performance of ANN and GANN algorithms.

\begin{figure}[h]
\centering
\includegraphics[width=0.45\columnwidth]{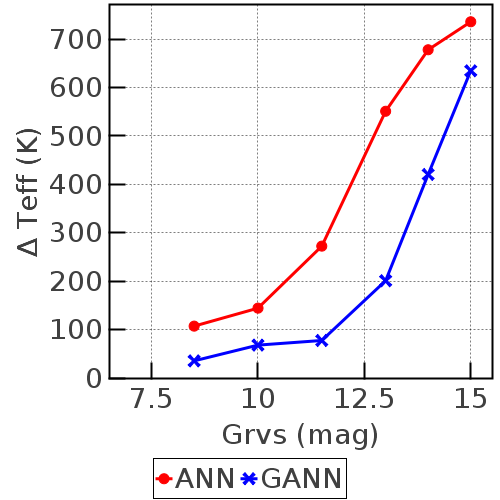}
\includegraphics[width=0.45\columnwidth]{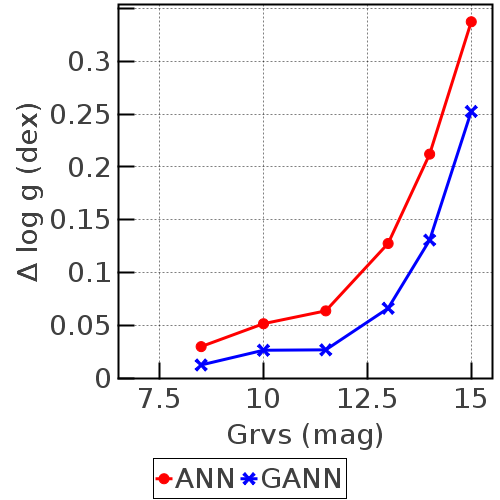}
\includegraphics[width=0.45\columnwidth]{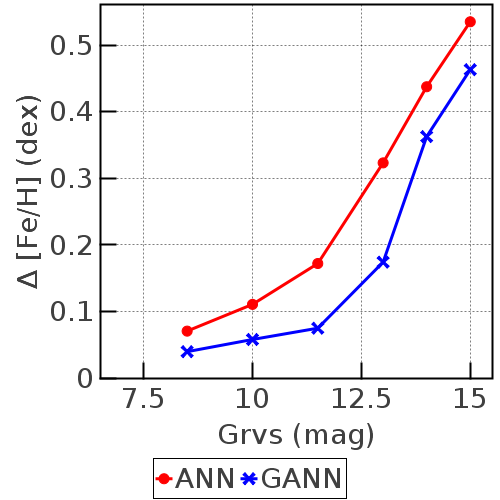}
\caption{68th percentile of residuals obtained by both ANNs and GANNs for early type stars sample: $T_{eff} \in \lbrack 7500, 11500 \rbrack$, $log\:g \in \lbrack 2, 5 \rbrack$, and $\lbrack Fe/H \rbrack \in \lbrack -2.5, 0.5 \rbrack$.}
\label{fig:earlyresults}
\end{figure}

\begin{figure}[h]
\centering
\includegraphics[width=0.45\columnwidth]{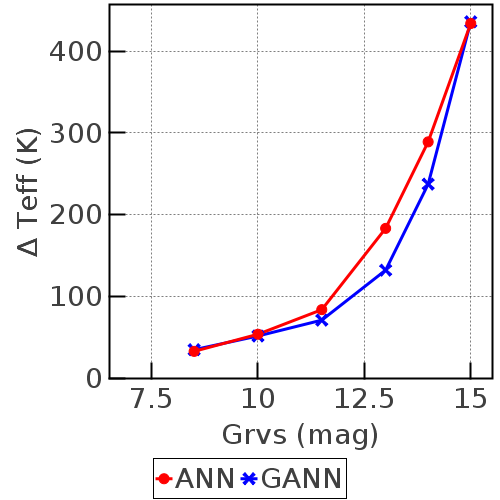}
\includegraphics[width=0.45\columnwidth]{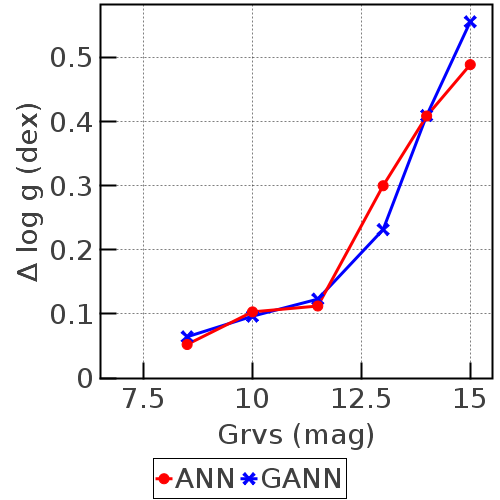}
\includegraphics[width=0.45\columnwidth]{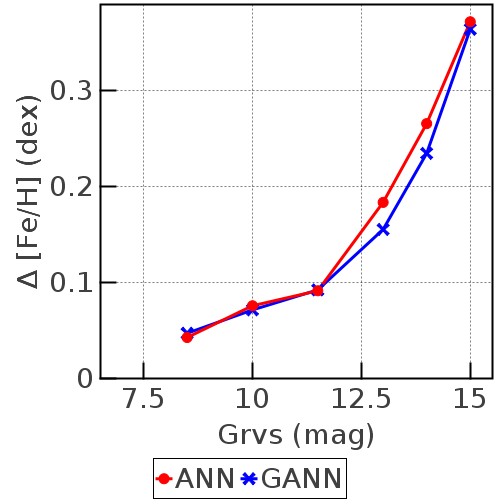}
\includegraphics[width=0.45\columnwidth]{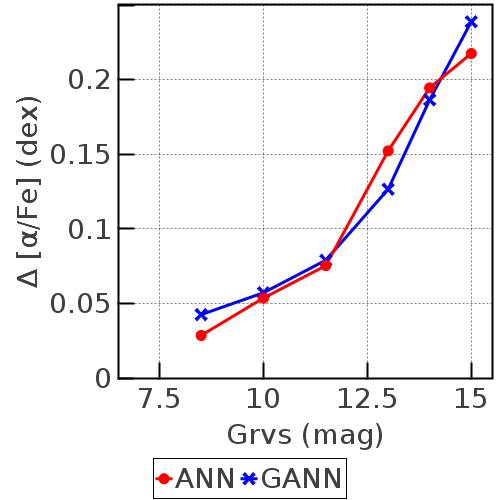}
\caption{68th percentile of residuals obtained by both ANNs and GANNs for late type stars sample: $T_{eff} \in \lbrack 4000, 8000 \rbrack$, $log\:g \in \lbrack 2, 5 \rbrack$, $\lbrack Fe/H \rbrack \in \lbrack -2.5, 0.5 \rbrack$, and $\lbrack \alpha/Fe \rbrack \in \lbrack -0.4, 0.8 \rbrack$.}
\label{fig:lateresults}
\end{figure}

\begin{figure*}
\centering
\begin{subfigure}{\columnwidth}
\captionsetup{width=0.97\columnwidth}
\centering
\includegraphics[width=0.45\columnwidth]{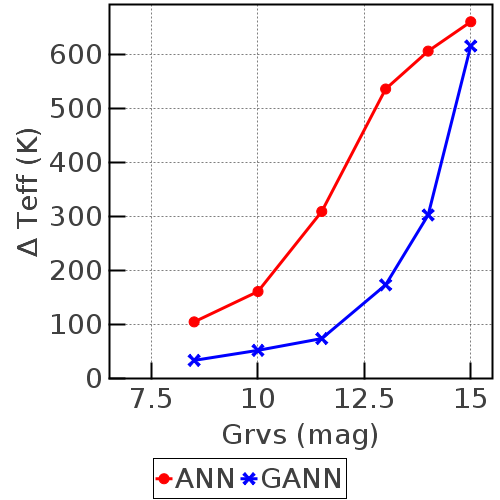}
\includegraphics[width=0.45\columnwidth]{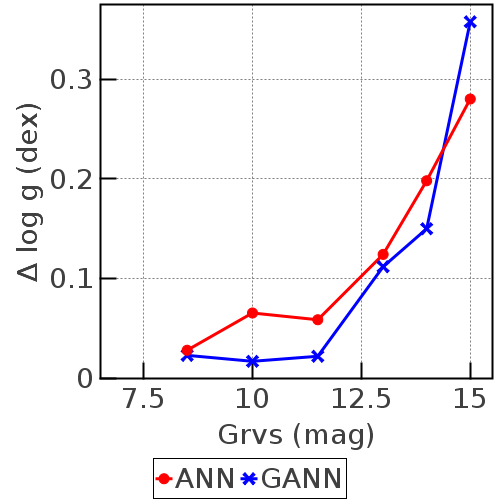}
\includegraphics[width=0.45\columnwidth]{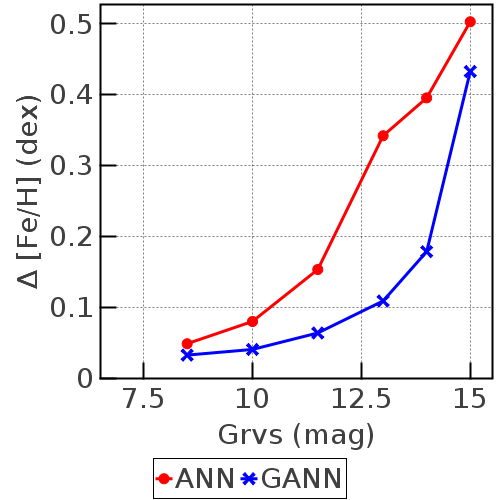}
\caption{Metal-rich giant stars: $T_{eff} \in \lbrack 7500, 11500 \rbrack$, $log\:g \in \lbrack 2.5, 3.5 \rbrack$, and $\lbrack Fe/H \rbrack \in \lbrack -0.5, 0.25 \rbrack$.}

\end{subfigure}
\begin{subfigure}{\columnwidth}
\captionsetup{width=0.97\columnwidth}
\centering
\includegraphics[width=0.45\columnwidth]{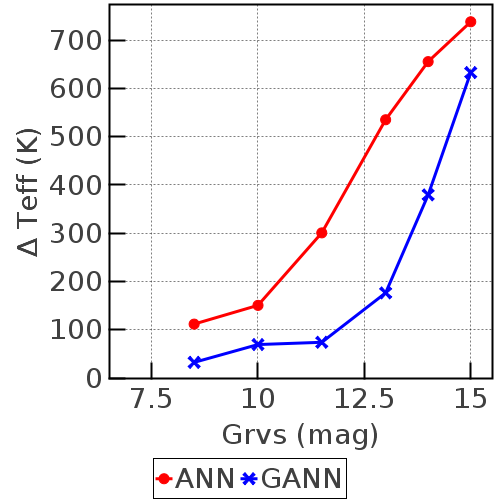}
\includegraphics[width=0.45\columnwidth]{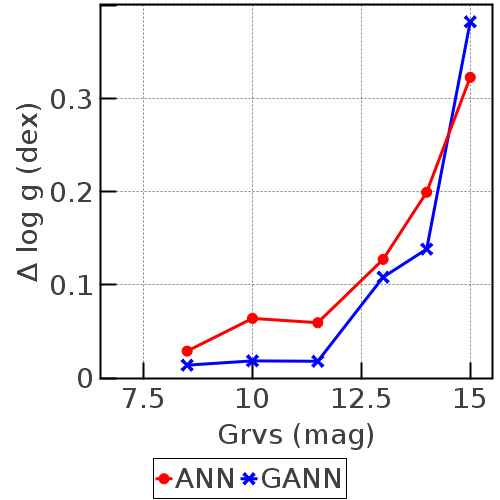}
\includegraphics[width=0.45\columnwidth]{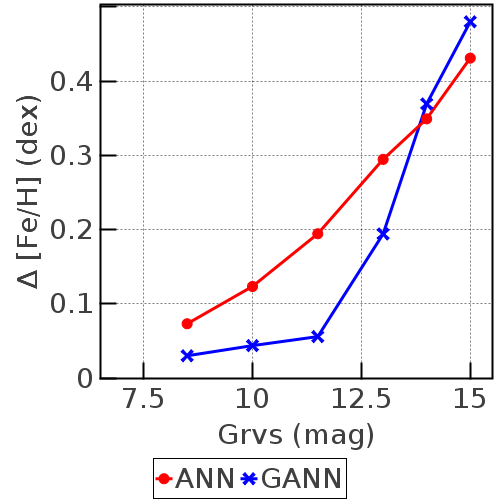}
\caption{Intermediate metallicity giant stars: $T_{eff} \in \lbrack 7500, 11500 \rbrack$, $log\:g \in \lbrack 2.5, 3.5 \rbrack$, and $\lbrack Fe/H \rbrack \in \lbrack -1.25, -0.5 \rbrack$.}

\end{subfigure}
\begin{subfigure}{\columnwidth}
\captionsetup{width=0.97\columnwidth}
\centering
\includegraphics[width=0.45\columnwidth]{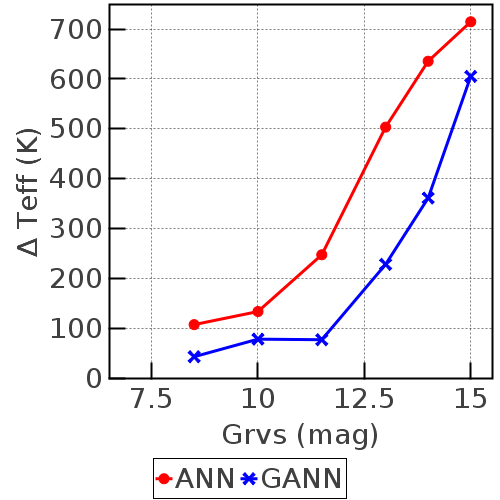}
\includegraphics[width=0.45\columnwidth]{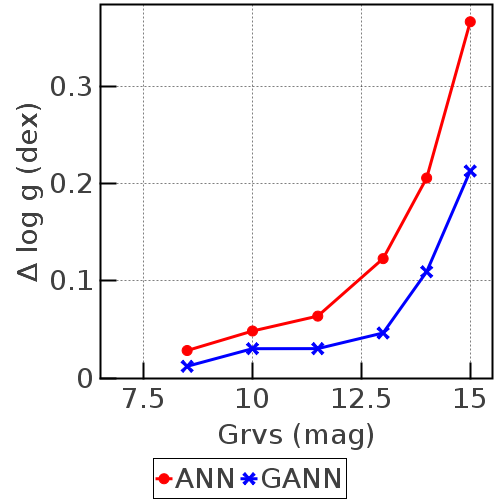}
\includegraphics[width=0.45\columnwidth]{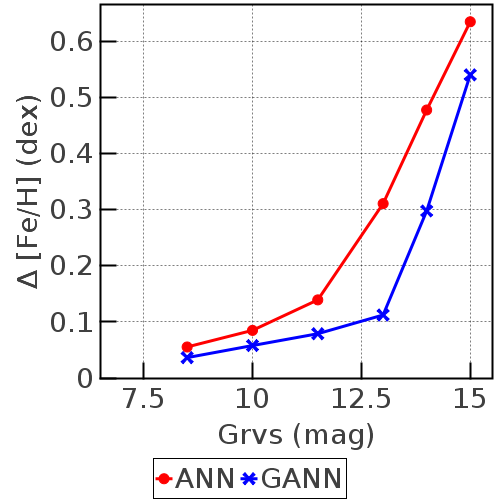}
\caption{Metal-rich dwarf stars: $T_{eff} \in \lbrack 7500, 11500 \rbrack$, $log\:g \in \lbrack 3.5, 4.5 \rbrack$, and $\lbrack Fe/H \rbrack \in \lbrack -0.5, 0.25 \rbrack$.}

\end{subfigure}
\begin{subfigure}{\columnwidth}
\captionsetup{width=0.97\columnwidth}
\centering
\includegraphics[width=0.45\columnwidth]{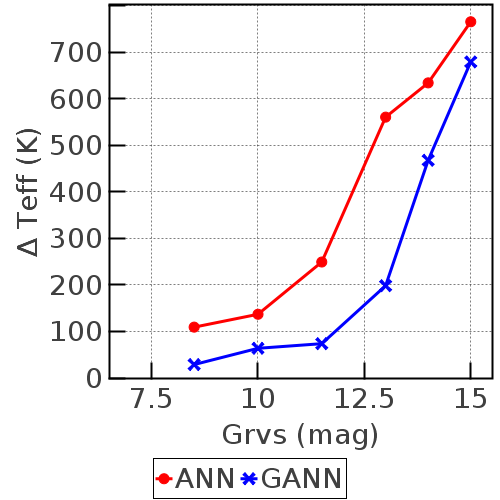}
\includegraphics[width=0.45\columnwidth]{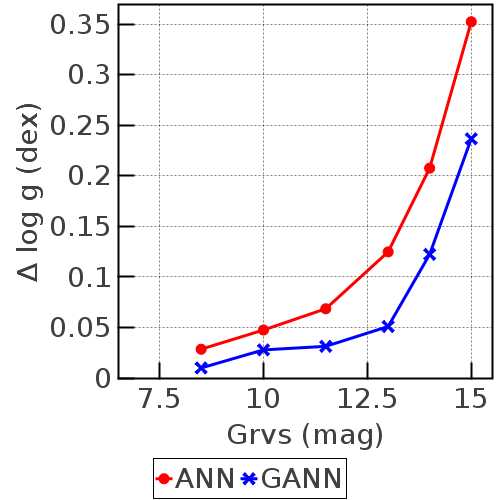}
\includegraphics[width=0.45\columnwidth]{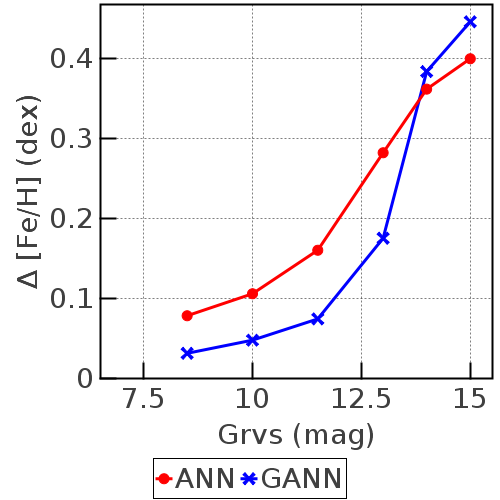}
\caption{Intermediate metallicity dwarf stars: $T_{eff} \in \lbrack 7500, 11500 \rbrack$, $log\:g \in \lbrack 3.5, 4.5 \rbrack$, and $\lbrack Fe/H \rbrack \in \lbrack -1.25, -0.5 \rbrack$.}

\end{subfigure}
\caption{68th percentile of residuals obtained for different early type star populations by both ANNs and GANNs.}
\label{fig:early_populations}
\end{figure*}

A first step to evaluate the performance of the algorithms is to show their behaviour as a function of $G_{rvs}$ magnitude (or, equivalently, of S/N). This is shown in Figures \ref{fig:earlyresults} and \ref{fig:lateresults}. We display the residuals obtained when the test dataset is presented to the ANNs and to the GANNs, for early and late type star samples, respectively, when working in the maximum likelihood mode of the algorithm. Following RB2016, the 68th quantile (Q68) of every AP is given to summarize the residuals obtained.

The parameterization errors found by ANNs are, in general terms, similar to those presented in RB2016, although a better performance is achieved in some parameters and for some particular types of stars. This point will be briefly commented hereinafter. Regarding the behaviour of GANNs as compared to ANNs, an improved parameter derivation for early stars can clearly be observed. In general terms, we also find an enhancement in the GANNs performance for late stars, but it is much more modest.

\begin{table}[h]
\begin{center}
\begin{tabular}{lll} \hline
\textbf{Parameter} & \textbf{Population} & \textbf{Specification}\\
\hline
\multirow{3}{*}{Metallicity ($\lbrack Fe/H \rbrack$)} & Poor & $\lbrack -2.5, -1.25 )$\\
 & Intermediate & $\lbrack -1.25, -0.5)$\\
 & Rich & $\lbrack -0.5, 0.25 \rbrack$\\
\hline
\multirow{2}{*}{Surface gravity ($log\:g$)} & Giant & $\lbrack 2.5, 3.5 )$\\
 & Dwarf & $\lbrack 3.5, 4.5 \rbrack$\\
\hline
\end{tabular}
\end{center}
\caption{Specifications of the different star populations taken into account.}
\label{tab:populations}
\end{table}

\begin{figure*}
\centering
\begin{subfigure}{\columnwidth}
\captionsetup{width=0.97\columnwidth}
\centering
\includegraphics[width=0.45\columnwidth]{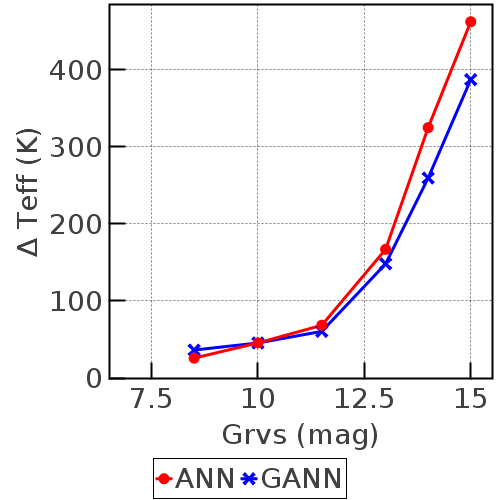}
\includegraphics[width=0.45\columnwidth]{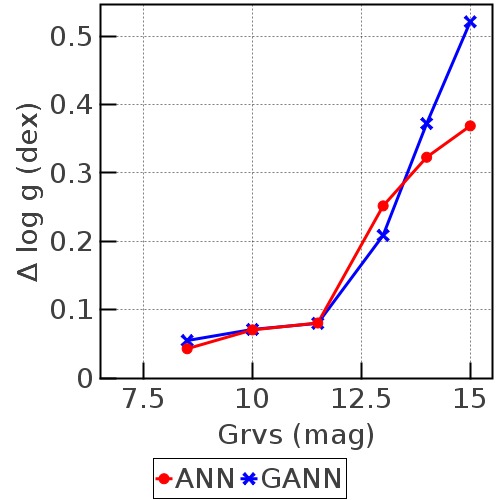}
\includegraphics[width=0.45\columnwidth]{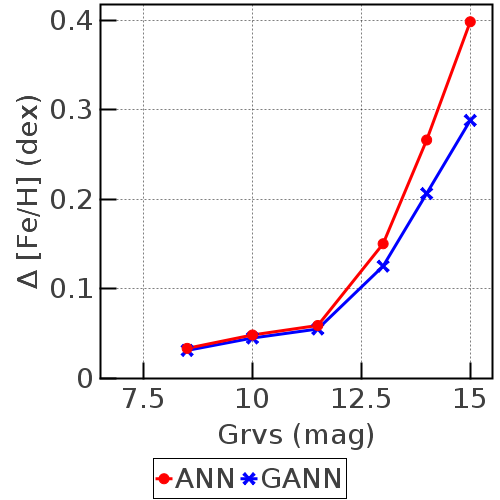}
\includegraphics[width=0.45\columnwidth]{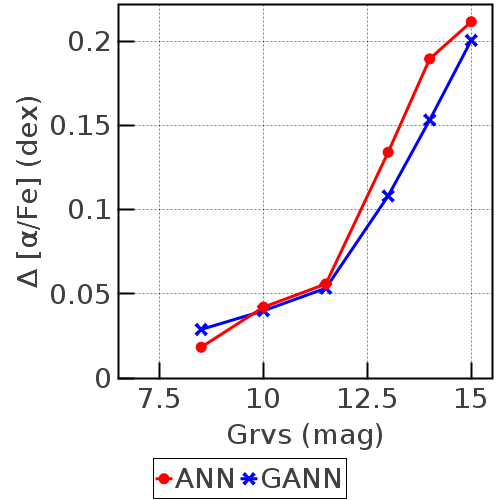}
\caption{Metal-rich dwarf stars: $T_{eff} \in \lbrack 4000, 8000 \rbrack$, $log\:g \in \lbrack 3.5, 4.5 \rbrack$, $\lbrack Fe/H \rbrack \in \lbrack -0.5, 0.25 \rbrack$, and $\lbrack \alpha/Fe \rbrack \in \lbrack -0.4, 0.8 \rbrack$.}

\end{subfigure}
\begin{subfigure}{\columnwidth}
\captionsetup{width=0.97\columnwidth}
\centering
\includegraphics[width=0.45\columnwidth]{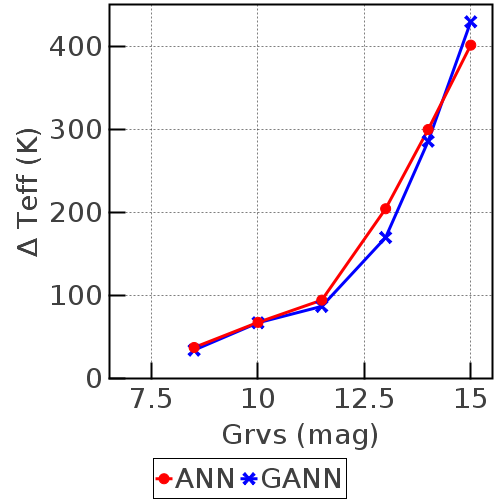}
\includegraphics[width=0.45\columnwidth]{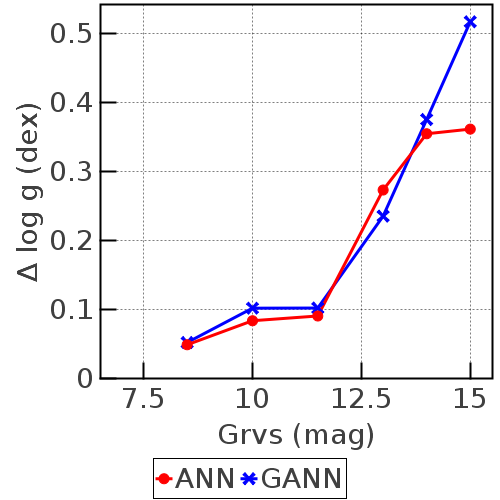}
\includegraphics[width=0.45\columnwidth]{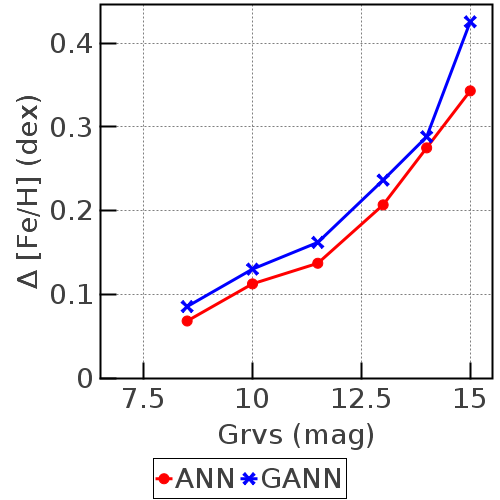}
\includegraphics[width=0.45\columnwidth]{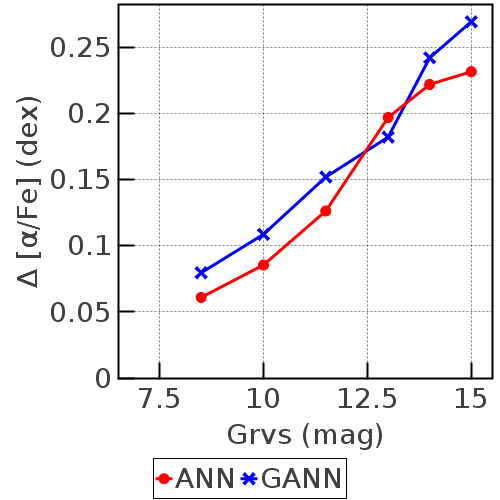}
\caption{Metal-poor dwarf stars: $T_{eff} \in \lbrack 4000, 8000 \rbrack$, $log\:g \in \lbrack 3.5, 4.5 \rbrack$, $\lbrack Fe/H \rbrack \in \lbrack -2.25, -1.25 \rbrack$, and $\lbrack \alpha/Fe \rbrack \in \lbrack -0.4, 0.8 \rbrack$.}
\end{subfigure}
\caption{68th percentile of residuals obtained for different late type star populations by both ANNs and GANNs.}
\label{fig:late_populations}
\end{figure*}

A second step is to evaluate the performance for different types of stars. This has been done by consideration of different typical Galactic stellar populations, whose specifications are shown in Table \ref{tab:populations}. With this aim we adopted the definitions for metallicity, gravity and effective temperature ranges included in RB2016. In Figures \ref{fig:early_populations} and \ref{fig:late_populations} we show the results for a selection of stellar types. Figure \ref{fig:early_populations} displays an example of the performance of our algorithms for the early star sample when parameterizing giants and dwarfs, with rich or intermediate metal content. In general terms, GANNs better parameterize all the parameters for the complete brightness range, and the results for temperature are slightly better for giants than for dwarfs. Metallicity residuals as low as $0.1$ dex for stars with $G_{rvs}<12.5$ magnitudes will certainly allow metallicity studies for these disc stars well outside the solar neighbourhood. Figure \ref{fig:late_populations} shows some results obtained for our late dwarfs sample, this time with rich and poor metal content. Consistently both $\lbrack Fe/H \rbrack$ and $\lbrack \alpha/Fe \rbrack$ are better parameterized for metal-rich than for metal-poor dwarfs, while the residuals in $T_{eff}$ and  $log\:g$ are very similar in both cases.

\begin{figure}[h]
\centering
\begin{subfigure}{\columnwidth}
\centering
\includegraphics[width=0.45\columnwidth]{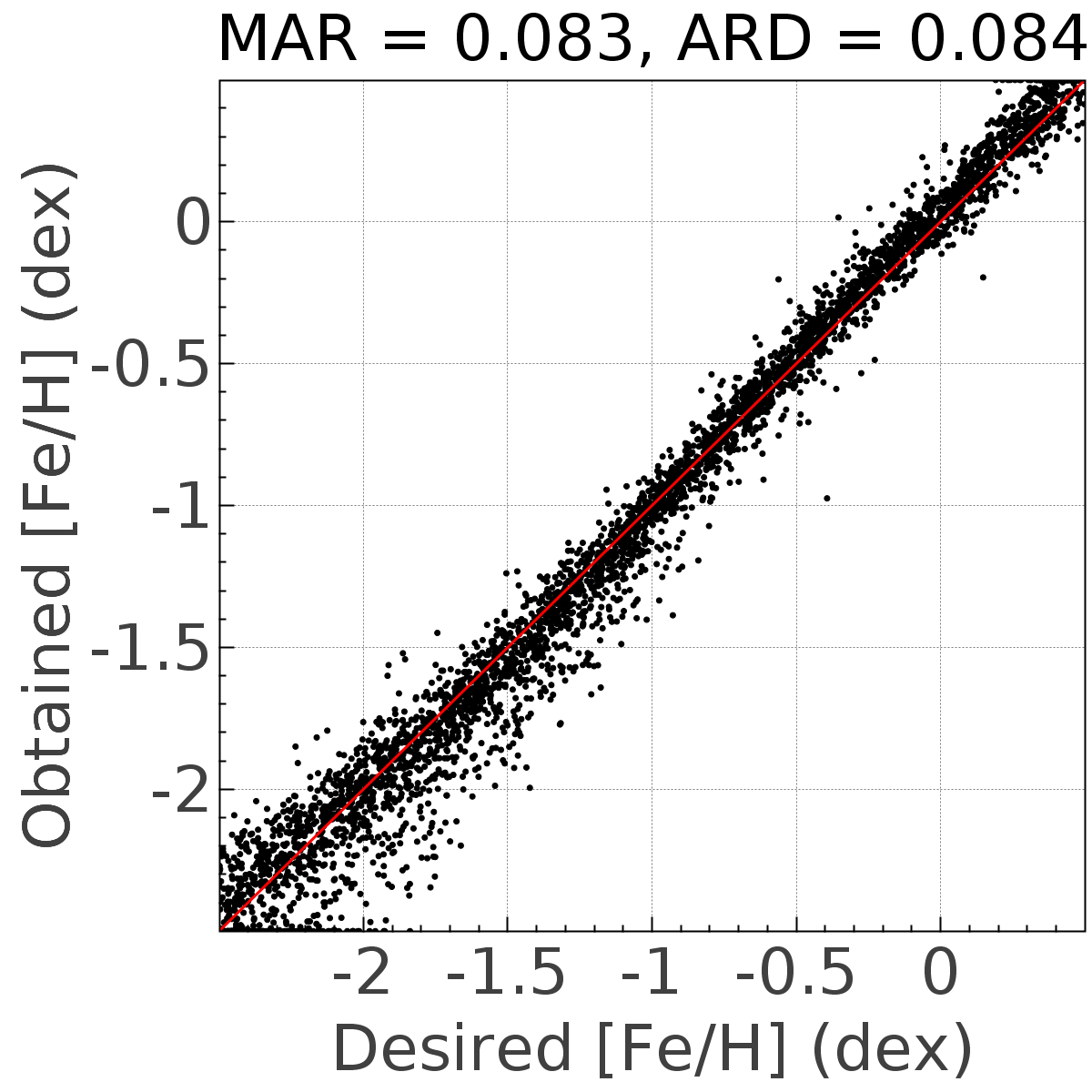}
\includegraphics[width=0.45\columnwidth]{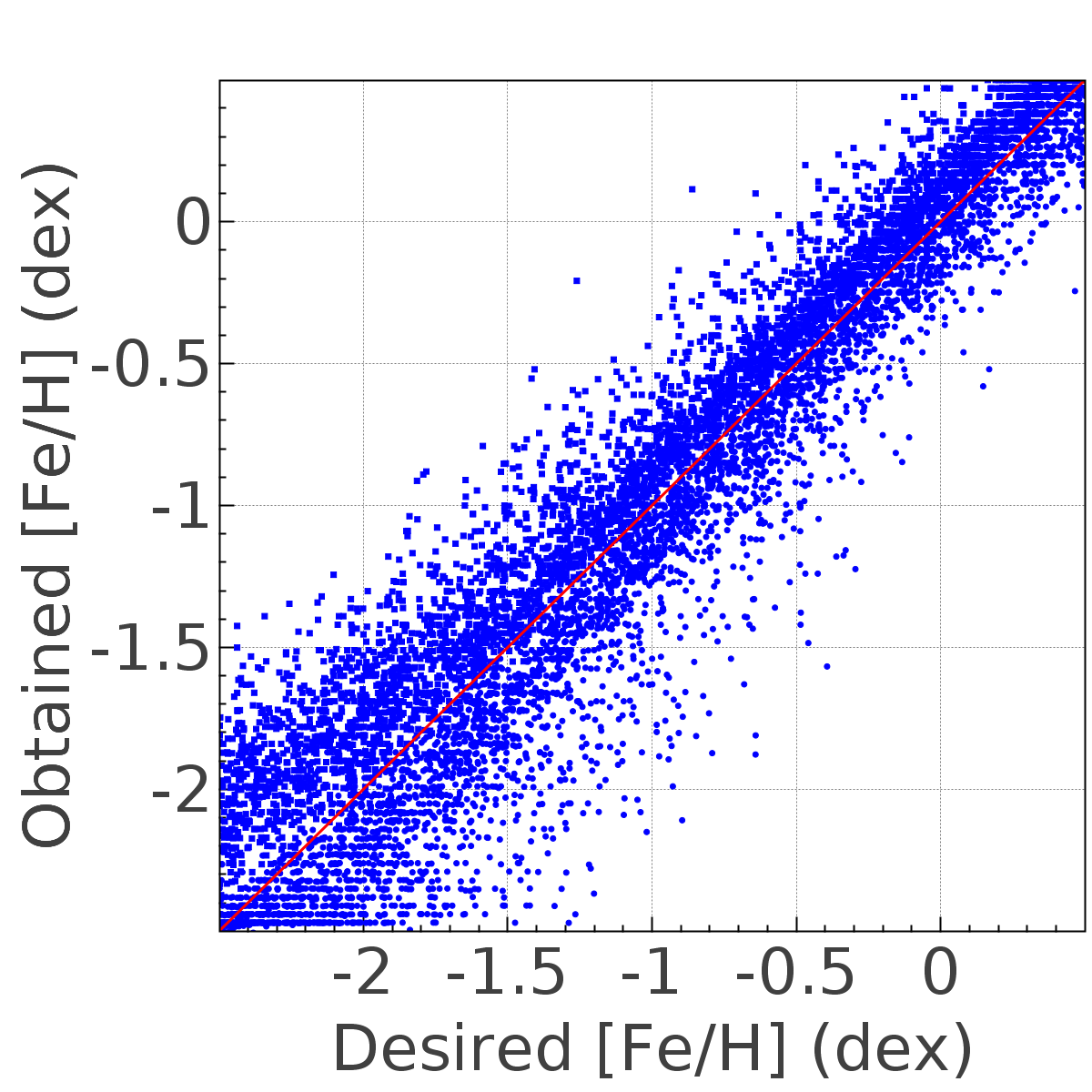}
\caption{$G_{rvs} = 8.5$ mag}

\end{subfigure}
\begin{subfigure}{\columnwidth}
\centering
\includegraphics[width=0.45\columnwidth]{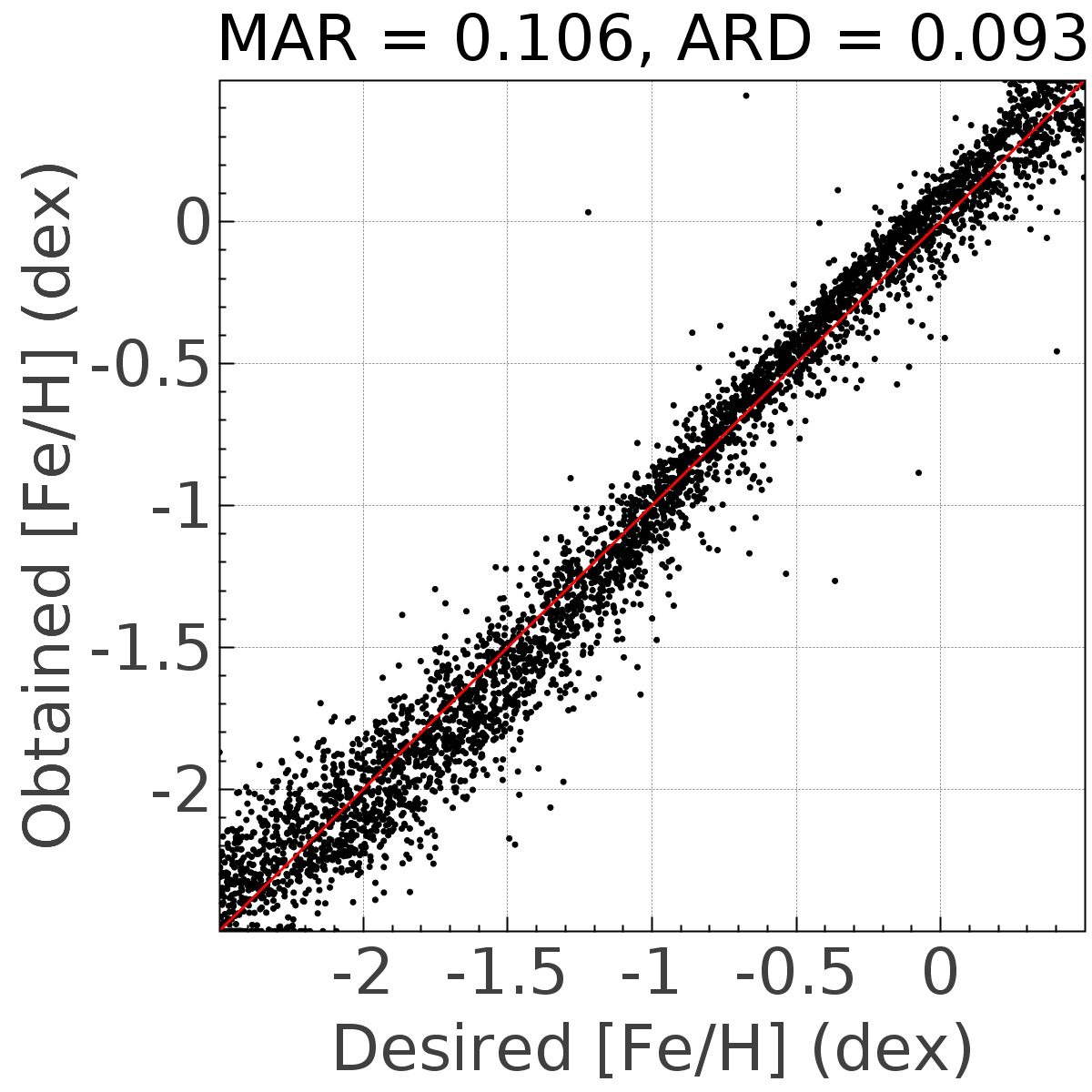}
\includegraphics[width=0.45\columnwidth]{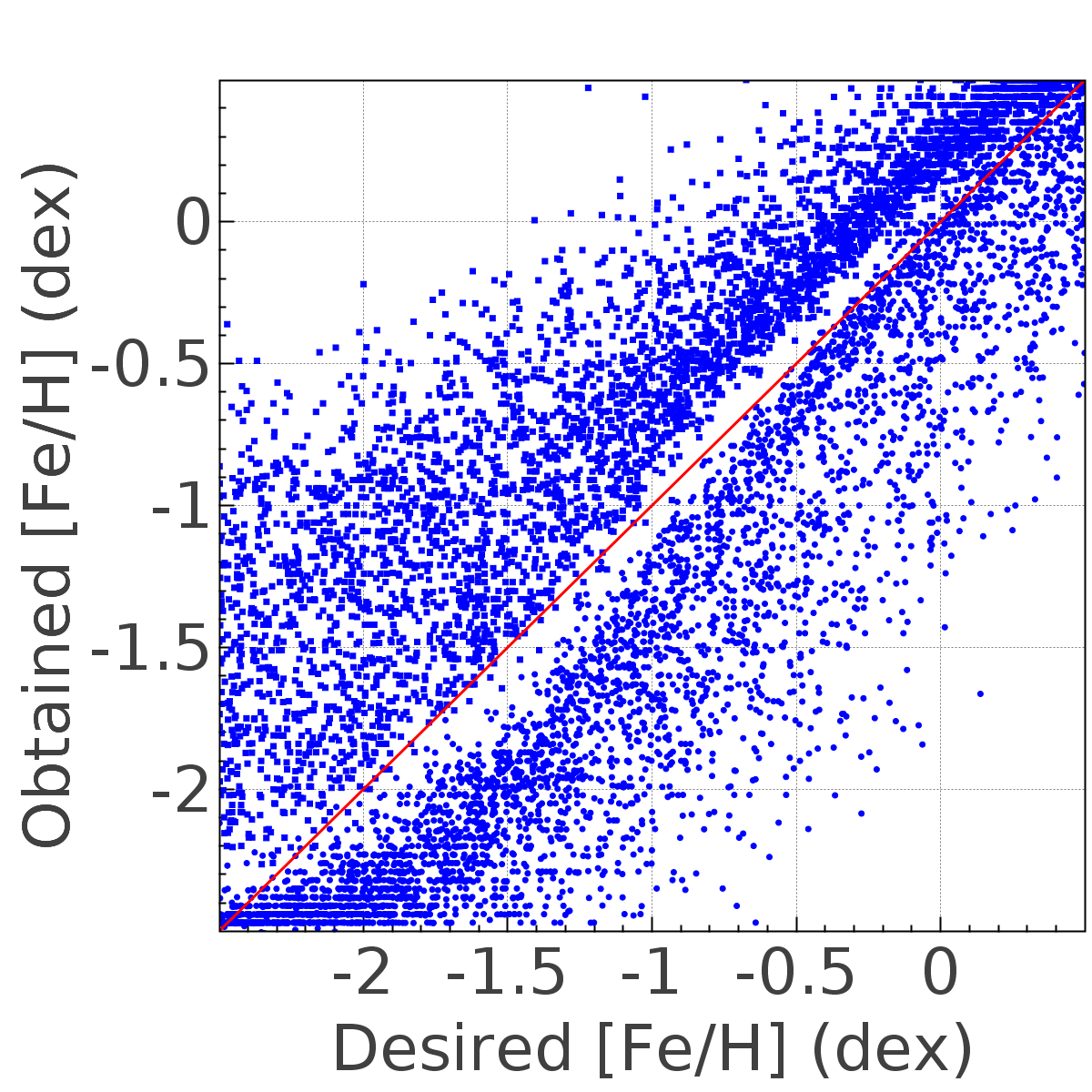}
\caption{$G_{rvs} = 10$ mag}
\end{subfigure}
\caption{Most probable estimation (left) and CIs at a level of confidence of $70\%$ (right) on $\lbrack Fe / H \rbrack$ for each observation from the early stars sample. Values of the mean and deviation of the fitting are also shown.}
\label{fig:met_cis}
\end{figure}

\begin{figure}[h]
\centering
\begin{subfigure}{\columnwidth}
\centering
\includegraphics[width=0.45\columnwidth]{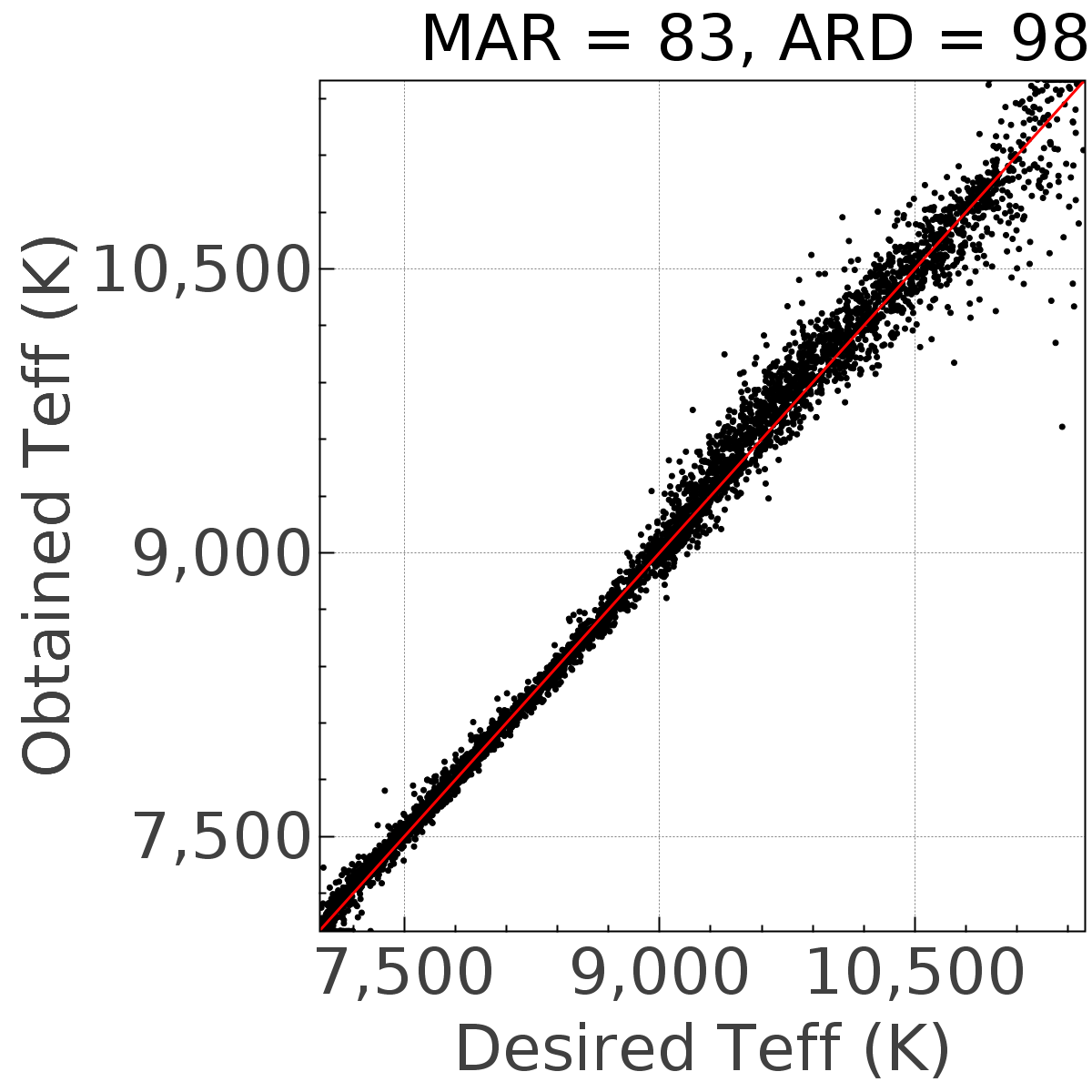}
\includegraphics[width=0.45\columnwidth]{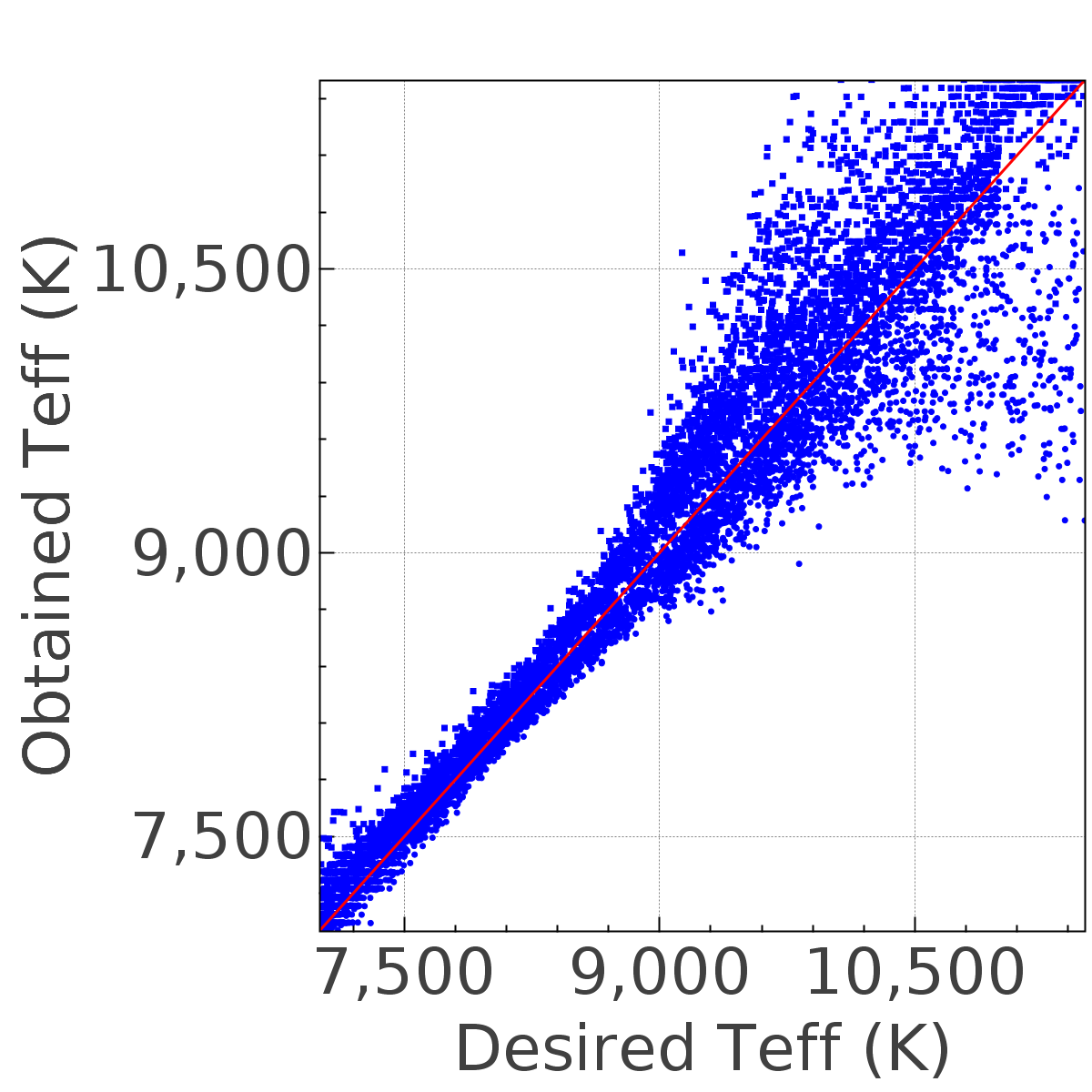}
\caption{$G_{rvs} = 8.5$ mag}
\end{subfigure}
\begin{subfigure}{\columnwidth}
\centering
\includegraphics[width=0.45\columnwidth]{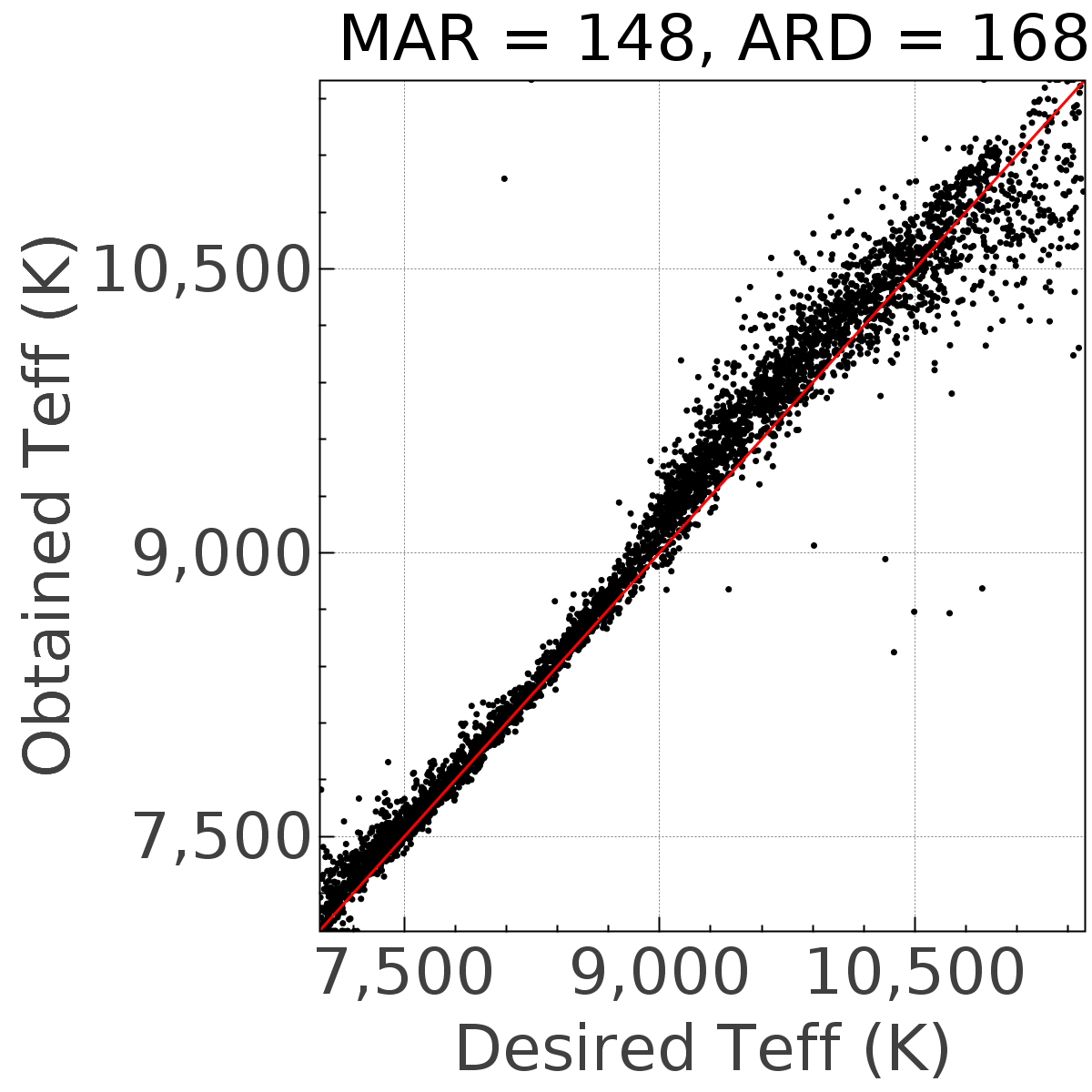}
\includegraphics[width=0.45\columnwidth]{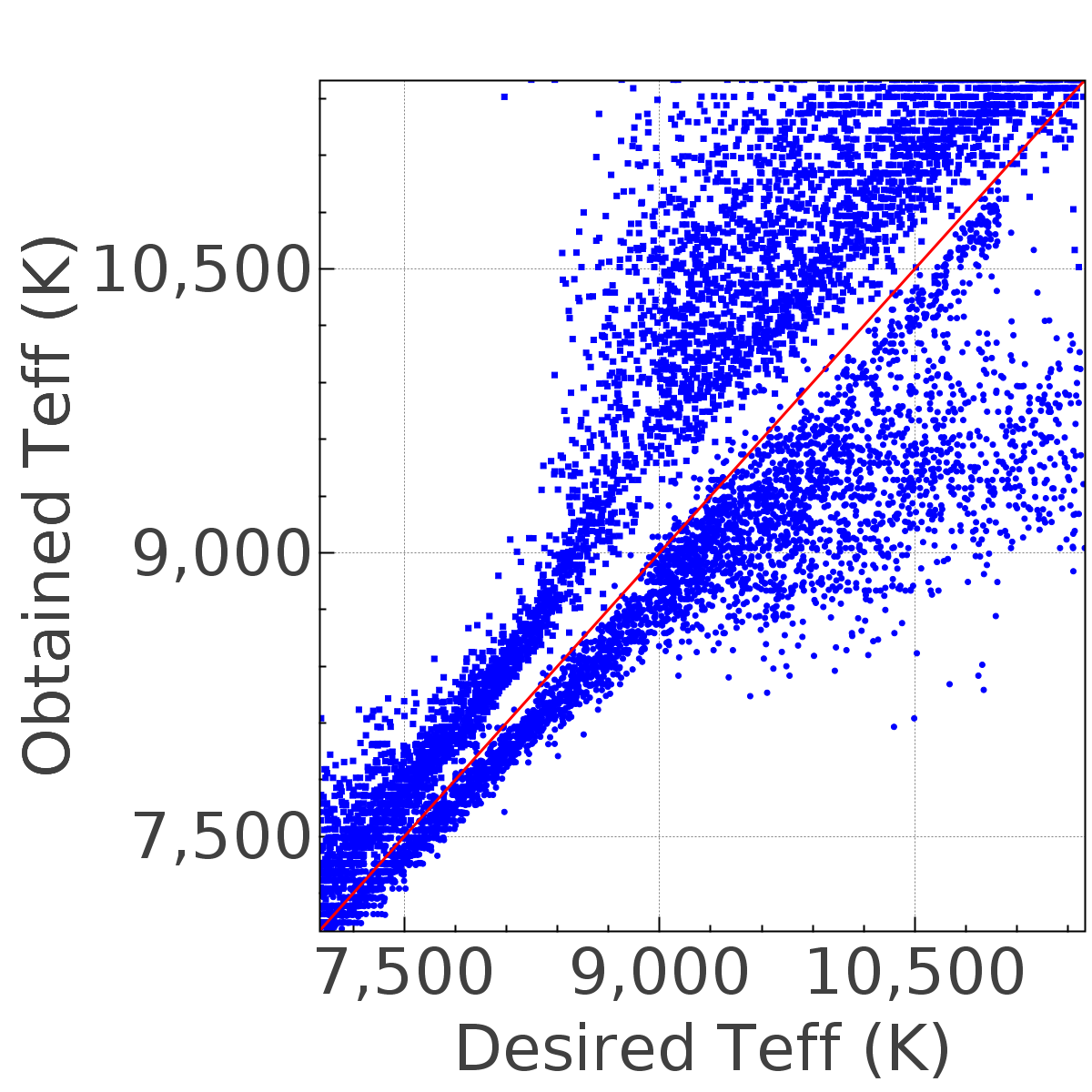}
\caption{$G_{rvs} = 10$ mag}
\end{subfigure}
\caption{Most probable estimation (left) and CIs at a level of confidence of $70\%$ (right) on $T_{eff}$ for each observation from the early stars sample. Values of the mean and deviation of the fitting are also shown.}
\label{fig:tef_cis}
\end{figure}

From Figure 5 we can conclude that the ANN algorithm residuals for a $13th$ magnitude late type star is around $210$ K in $T_{eff}$, $0.32$ dex in $log\:g$, $0.20$ dex in $\lbrack Fe/H \rbrack$, and $0.175$ in $\lbrack \alpha/Fe \rbrack$, confirming a better parameterization by ANN for low S/N spectra as presented in RB2016 (see for instance Figure 9 in that paper). It is also remarkable that in most cases, the errors reported by the use of GANNs even improve these values for the faintest stars in the sample.

The fully Bayesian implementation of the method is illustrated in Figures \ref{fig:met_cis} and \ref{fig:tef_cis}. These figures show the most probable estimation of $\lbrack Fe/H \rbrack$ and $T_{eff}$ and the upper and lower confidence intervals (CIs) for each observation from the early stars sample using GANNs in the Bayesian mode. The level of confidence is $70\%$ and the results are shown for stars at two magnitude levels, $G_{rvs}= 8.5$ and $10$ mag. For each spectrum in the dataset, the parameterization produced by the ANN and 49 pseudo-random generated parameterizations are mutated over 10 steps using PSO algorithm, so that a full posterior distribution can be estimated using such parameterizations. Afterwards, all the APs posterior distributions are marginalized for each parameter, computing the corresponding CIs. Confidence intervals that would  encompass the true population parameter with a probability of $70\%$ are then calculated. Metallicity and effective temperature values outside such CIs are not statistically significant to the $30\%$ level under the assumptions of the experiment.

\begin{figure}[h]
\centering
\begin{subfigure}{\columnwidth}
\centering
\includegraphics[width=0.8\columnwidth]{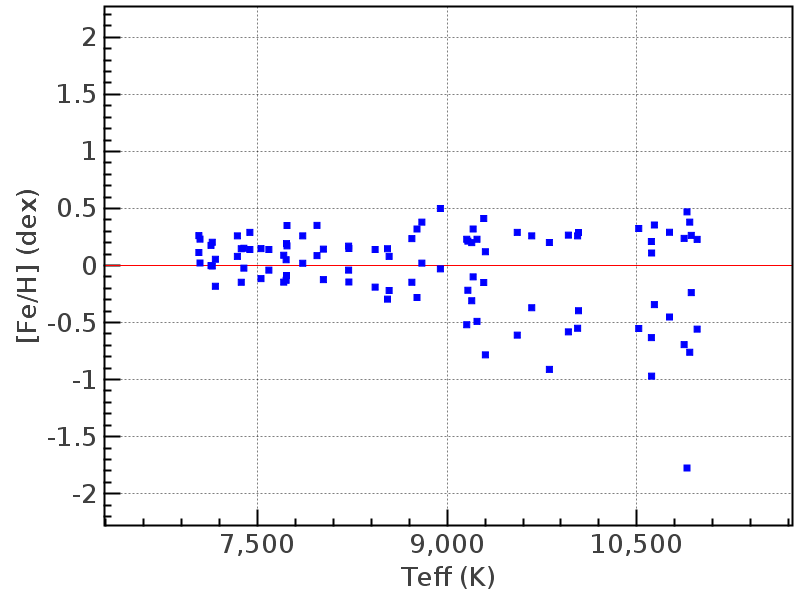}
\caption{$\lbrack Fe/H \rbrack$ as a function of $T_{eff}$ for stars with $log\:g=3.5$ dex.}
\end{subfigure}
\begin{subfigure}{\columnwidth}
\centering
\includegraphics[width=0.8\columnwidth]{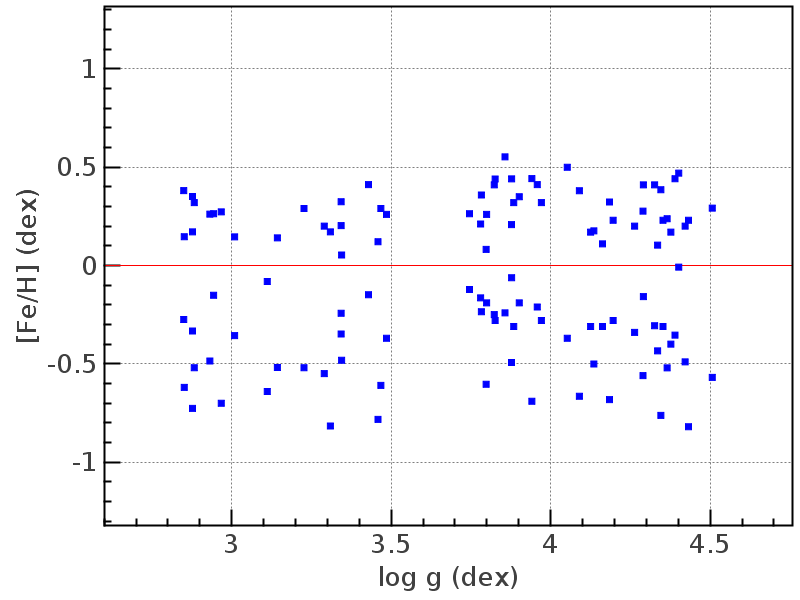}
\caption{$\lbrack Fe/H \rbrack$ as a function of $log\:g$ for stars with $T_{eff}=9500$ K.}
\end{subfigure}
\caption[x]{Confidence intervals at a confidence level of $70\%$ on $\lbrack Fe/H \rbrack$ for solar metallicity stars with $G_{rvs}=10$ mag from the early stars sample.}
\label{fig:confidenceintervals3}
\end{figure}

It can be observed that the CIs become broader with increasing $G_{rvs}$, since the residuals also increase with the stellar magnitude. Looking at the figures we can say that, in general terms, CIs obtained by GANNs are robust estimators of the residuals given by them. Figure \ref{fig:confidenceintervals3} shows the amplitudes of the metallicity confidence intervals for solar metallicity stars\footnote{For solar metallicity, we selected those stars among our random spectra with metallicities in the range $\lbrack -0.15, 0.15 \rbrack$} with $G_{rvs}=10$ mag as a function of $T_{eff}$ in the case of stars with $log\:g=3.5$ dex, and as a function of $log\:g$ for stars with $T_{eff}=9500$ K, at a confidence level  of $70\%$. This figure illustrates the fact that the derivation of $\lbrack Fe/H \rbrack$ is less accurate for hot stars due to the scarcity of metallic lines in their spectra. We also note that the CIs are not symmetric. This is due to the non-parametric computation of the posterior distributions, since it does not force the posteriors to be Gaussian or any predetermined distribution.

\section{Discussion\label{sect:discussion}}

ANNs are a great tool that offer nonlinear regression capabilities to any degree of complexity. Furthermore, they can provide accurate predictions when new data is presented to them, since they can generalize their solutions. However, in principle, they are not able to give a measure of uncertainty over their predictions. Giving a measure of uncertainty over predictions is desirable in application domains where posterior inferences need to assess the quality of the predictions, specially when the behaviour of the system is not completely known. This is the case for data analysis coming from complex scientific missions such as the Gaia satellite.

This work has presented a new architecture for ANNs, Generative ANNs (GANNs), that models the forward function instead of the inverse one. The advantage of forward modelling is that it estimates the actual observation, so that the fitness between the estimated and the actual observation can be assessed, which allows for novelty detection, model evaluation and active learning. Furthermore, these networks can be integrated in a Bayesian framework, which allows us to estimate the full posterior distribution over the parameters of interest, to perform model comparisons, and so on. However, GANNs require more computations, since the network needs to be evaluated iteratively before it reaches the best fit for the current observation. Some shortcuts that could reduce the number of required evaluations have been described, such as MCMC methods. In any case, the computation of AP uncertainties, taking into account all involved sources of errors, is possible with GANNs, while it is not feasible with other methods that use the Hessian matrix, at least without a significant implementation effort.  

The capability of both ANNs and GANNs to perform AP estimation was demonstrated by means of Gaia RVS spectra simulations, since they can efficiently contribute to the optimization of the parameterization. In most stellar types except metal poor stars, the parameterization accuracies are on the order of 0.1 dex in $\lbrack Fe/H \rbrack$ and $\lbrack \alpha/Fe \rbrack$ for stars with $G_{rvs}<12$, which accounts for a number in the order of four million stars.

Internal errors here reported will need to be combined in the near future with the external uncertainties that will be obtained when real spectra from benchmark reference stars are analysed. GANNs give significantly better AP estimations than ANNs for A, B and early F stars, and only marginally improved for cooler stars. Nevertheless, its use in the regular Gaia pipeline of data analysis has been postponed to the next development cycle due to its high computational cost.

The methodology presented here is not only valid for AP estimation, but is a general scheme that can be extrapolated to other application domains.

\begin{acknowledgements}
\\
This work was supported by the Spanish \emph{FEDER} through Grants ESP-2014-55996-C2-2-R, and AYA2015-71820-REDT.
\\
AU acknowledges partial financial support from the Spanish MECD, under grant PRX15/0051, to entitle a research visit to the Astronomy Unit, Queen Mary University of London, Mile End Road, London E1 4NS, UK.
\end{acknowledgements}

\bibliographystyle{aa}
\bibliography{aaGANN}

\end{document}